# Silicon Avalanche Pixel Sensor for High Precision Tracking

N. D'Ascenzo,[a] P.S. Marrocchesi,[b] C.S. Moon,[c,d] F. Morsani,[c] L.Ratti,[e]
V. Saveliev,[a,*] A. Savoy Navarro,[c,d] Q. Xie,[f]

[a] *Institute of Applied Mathematics, Russian Academy of Sciences, Moscow, Russia*
[b] *Univ. di Siena and INFN Pisa, Italy*
[c] *INFN Sezione di Pisa, Pisa, Italy*
[d] *Laboratoire APC, Université Paris-Diderot/CNRS, Paris, France*
[e] *INFN Sezione di Pavia, Pavia, Italy*
[f] *HUST, Wuhan, China*

*E-mail*: saveliev@mail.desy.de

ABSTRACT: The development of an innovative position sensitive pixelated sensor to detect and measure with high precision the coordinates of the ionizing particles is proposed. The silicon avalanche pixel sensors (APiX) is based on the vertical integration of avalanche pixels connected in pairs and operated in coincidence in fully digital mode and with the processing electronics embedded on the chip. The APiX sensor addresses the need to minimize the material budget and related multiple scattering effects in tracking systems requiring a high spatial resolution in the presence of a large occupancy. The expected operation of the new sensor features: low noise, low power consumption and suitable radiation tolerance. The APiX device provides on-chip digital information on the position of the coordinate of the impinging charged particle and can be seen as the building block of a modular system of pixelated arrays, implementing a sparsified readout. The technological challenges are the 3D integration of the device under CMOS processes and integration of processing electronics.

---

[*] Corresponding author.

# Contents

## 1. Introduction

Position sensitive detectors are the building blocks of charged particle tracking systems. Over the years, the latter have been evolving in design and complexity, fostering new ideas in particle detection and often leading to the development of innovative sensors. The main requirements for a tracking system (e.g.: granularity, position accuracy, low material budget, rate capability, power consumption, radiation resistance, cost-effectiveness, simplicity in building and integration) may vary among different experiments. Nevertheless, a number of common issues have guided the development of cutting-edge technologies in the design of semi-conductor based detectors.

While, for example, multiple scattering may only be a minor issue for a tracking system in a space-borne Astroparticle Physics experiment, designed to detect a single particle of cosmic origin with a particle energy in the multi-TeV range, on the contrary the minimization of material budget is an important goal for most of the present (and future) multi-particle tracking systems at colliders experiments. Therefore, a considerable effort has been devoted to the development of low material budget position detectors.

In the following, we provide just a few examples of the stringent requirements that have been established in a number of HEP experiments with colliding beams at high luminosity, where precise vertex information, primarily extracted from high-resolution position measurements near the interaction point (IP) by a dedicated vertex detector, is crucial [1].

The measurement of time dependent CP asymmetries in $B^0$ decays is a good example. The BABAR Silicon Vertex Tracker (SVT) detector has reached these goals with a five-layered silicon strip detector [2].

The more recent SuperB SVT design [3] was based on the BABAR vertex detector layout with the addition of an innermost layer - closer to the IP and known as Layer0 - which was required to stand typical rates of 100 MHz/cm$^2$ and designed with a radius of about 1.5 cm, high granularity (50x50 microns), low material budget (close to 1% $X_0$), and adequate radiation resistance. The SVT innermost layer had to provide good space resolution, while coping with high background levels, which made a possible solution based upon a pixelated system the most appealing. However, keeping the material budget low enough not to deteriorate the tracking performance is challenging, and this generated a remarkable activity in the development of thin hybrid pixels and thin pixel sensors based on CMOS Monolithic Active Pixel Sensor (MAPS) technology [4]. A first option for pixelated detectors was based on hybrid pixel sensors with a data-push readout architecture (featuring data sparsification on pixel and timestamp information for the hits) that had been developed by the SLIM5 Collaboration [5]. A second option for the SuperB SVT was based on CMOS MAPS, whose main advantage with respect to hybrid pixels is that they can be very thin, as the sensor and the readout are incorporated in a single CMOS layer, only a few tens of micron thick. This resulted in a design where the deep N-well (DNW) of a triple well commercial CMOS process was used as charge collecting electrode and extended to cover a large fraction of the elementary cell. Use of a large area-collecting electrode allowed the designer to include also PMOS transistors in the front-end. The full analog signal processing chain implemented at the pixel level (charge preamplifier, shaper, discriminator and latch) could be partly realized in the p-well physically overlapped with the area of the sensitive element, allowing the development of a complex in-pixel logic with similar functionalities as the hybrid pixels.

Parallel to these important advances in charged particle detection, the development of a new photo-sensor known as Silicon Photomultiplier (SiPM) based on the avalanche structures has successfully been implemented [6]. Of particular interest in such devices are the pixelated structure, the high intrinsic amplification gain, its capability to detect low light intensities, down to a single photoelectron, compactness of the device, the insensitivity to strong magnetic fields and the relatively low bias voltage. The use of silicon avalanche structures to detect charged particles has been first proposed in [7], with an innovative concept, whereby the signals from two vertically aligned avalanche pixels structure are detected in coincidence upon the passage of a charged particle. Such structures overcome the main problems of detection of the charged particles by the SiPM, in particular the sensitivity to photons and effect of dark rate.

The present work investigates on the development of an Avalanche Pixel Sensor (APiX), based on the 3D-vertical integration of avalanche pixels and implementation of the dedicated fully digital readout electronics on the same structure. The large signal amplification provided by the controlled avalanche processes would generate enough charge (despite the small thickness of the layer and the inherent fluctuations of the ionization yield or energy straggling) to overcome the need of an analog front-end amplification stage. Signals above threshold would generate a digital signal upon the passage of a charged particle crossing the overlapping active areas on the two vertically aligned cells. Therefore, a position sensitive detector with digital information and sparsified readout of the position of the hit could be envisaged, featuring low power consumption, low noise and high detection efficiency. With a suitable design, an avalanche pixel detector, structured into elementary cells of a few tens of microns (as an example), can be manufactured, under standard CMOS processes and include the necessary electronics on the same substrate. The implementation of the device in standard CMOS processes is challenging, but within reach of the available technology.

## 2. Detector Concept

### 2.1 APiX Sensor Concept

The general structure of APiX is presented in Fig 1. The avalanche pixel sensor consists of a substrate and a pair of vertically aligned avalanche cells, located on either layer, defining the single pixel element. On the first sensor layer, the avalanche sensor element operates in breakdown mode for detecting ionizing particle. A second avalanche sensor element, also operating in breakdown mode and situated on the second layer, will detect the ionizing particle seen by the first avalanche sensor element and experience substantially coincident breakdown with the first avalanche sensor element, in response to the ionizing particle traversing both elements. A logic element, electrically interconnected to the first and second avalanche sensors, will process the signal or signals created by the coincident breakdown of the first and second avalanche sensor elements, therefore allowing the signal (or signals) to be distinguished from a dark signal event in either the first or second sensor element and from the signals produced by background photons.

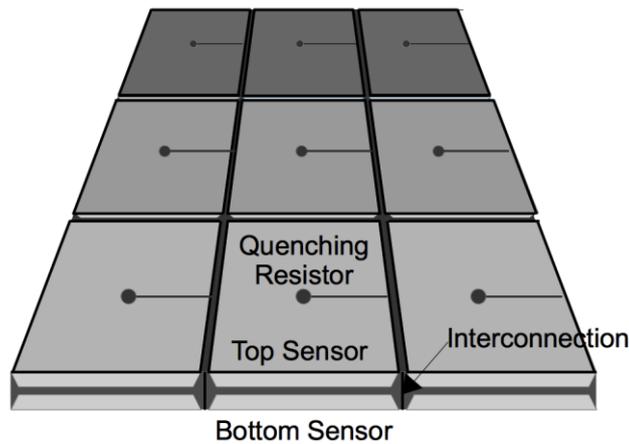

Fig.1 General view of APiX structure showing the vertically alignment
avalanche sensors with quenching elements and interconnection

The concept of a double-layered array based on this technology is then defined by pairing an avalanche sensor from the first layer to the corresponding vertically aligned avalanche sensor in the second layer. It provides a position sensitive avalanche pixel array, includes an interconnection of the first and second layer of the substrate, and logic elements for providing output signals to the readout electronics. The output signals contain the information corresponding to the number and position of sensor pairs activated during a charged particle flux exposure.

The avalanche structure is operated in breakdown mode, meaning that the avalanche current increases exponentially. The process is controlled by a high resistivity quenching element placed in series to each avalanche structure. As the current through the sensor elements increases, the voltage drop across the quenching resistor also increases until the voltage applied to the avalanche structure is below the breakdown voltage. The negative feedback mechanism introduced by the quenching resistor causes a quenching of the avalanche process and its termination. In this configuration, the avalanche structure reaches an intrinsic gain up to $10^6$. Quenching elements may be produced by CMOS technology, for reaching the high resistive values required for effective quenching. The thickness of the common substrate is small enough to allow the ionizing radiation to penetrate across the whole pixel and to interact with both avalanche sensors in the two vertically alignment layers of the APiX structure. Due to the high intrinsic gain of the avalanche sensors, the detected radiation in the APiX pixel corresponds to a pair of signals from the two opposite avalanche sensors whose amplitude are well above the electronics noise. The nature of these signals can be considered semi-digital, because the amplitude of the output signal is not defined by the number of carriers created in the sensitive volume, but defined by the avalanche process, triggered by ionizing particle. When an ionizing particle is detected by the APiX pixel the two signals generated in the two opposite avalanche sensors show only that these detector elements were both activated. The values of the signals are identical and independent of the energy deposited by the charged particle in the avalanche structures. It depends uniquely on the characteristic parameters of the avalanche structure.

Because of the nature of signals of avalanche sensors, a signal produced by the ionizing particle is not distinguishable from signals produced by thermally generated electron/hole pairs or produces by the photons. A key feature of the APiX structure, to overcome this problem, is the inclusion of logic elements in order to make fully use of the digital nature of the signal of the avalanche structures and the inclusion of an interconnection line between the two avalanche structures in the opposite surfaces of the pixel Fig.2.

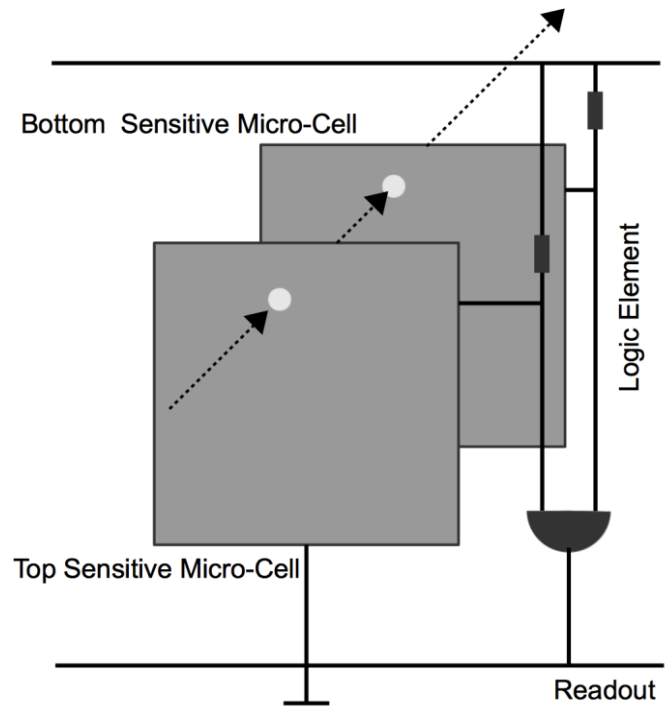

Fig. 2. Logical scheme of the APiX sensor.

The logic determines whether top and bottom avalanche structures in the APiX pixel are simultaneously activated, indicating the detection of a charged particle passing through the pixel. The dark rate due to the thermal electron/hole pairs is strongly suppressed in the coincidence, because the spontaneous breakdown is a stochastic process in each avalanche sensor. The same effect of suppression will work for the photons, which could be detected by the avalanche structures, but in this case the signal is generated only in one single avalanche sensor.

Another possible source of background events, with corresponding activation of the avalanche sensors, is the optical crosstalk between adjacent sensors. During the avalanche processes, optical photons are produced which can be detected in the nearby sensor, originating a signal indistinguishable from the detection of a charged particle. This effect can be overcome by introducing trenches filled with optically isolating material between adjacent sensors. Optical isolating elements include optically non-conductive material, such as polyimide, or metal capable of absorbing the optical photons. The required trenches are narrow and consume relative small amount of the pixel area, therefore affecting the geometrical efficiency of the detector to a limited extent.

The APiX structure is composed of an array of a few thousand of such pixels, with variable size and shape according to the application. The logic output of the pixel activation allows the determination of the position of the detected charged particle. In addition, the sum of the logic outputs of all pixels allows measuring the total number of detected charged particles, hence providing a measurement of the intensity of charged particles with very high sensitivity. The total thickness of the APiX sensor may be in the order of ten microns, including the electronics, making this detector competitive for low material budget applications.

## 2.2 APiX Electronics Issues

The design of the electronics, to be integrated in the sensor cell, is a crucial step, in that the detection efficiency of the sensor is critically dependent on the fill-factor. Therefore, the area covered by the electronics in the detector element has to be minimized, starting from the quenching network. A quenching resistor is the simplest choice, but might not be the most convenient solution in terms of silicon area. On the other hand, a network consisting of active devices, while having the potential for a significant area reduction, has to comply with the relatively large voltage drop set by the detector biasing requirements. It should be noted here that the quenching network, usually done by implementing resistors and critical in the photon detection is indeed not critical for our case, i.e. the ionizing particle detection, because it could be positioned on top of the sensors.

The very front-end circuit has to perform the task of converting the voltage pulse from the sensor directly to a digital level, compatible with the logic levels of the selected CMOS process. This stage may consist of a simple inverter or a memory cell. In the case of the inverter, the signal duration will depend on the characteristic discharge time of the sensor. In the case of the latch, signal duration can be controlled through a reset signal, resulting in a higher degree of flexibility in signal processing on the one hand, in a more complex logic control circuitry and routing on the other. Capacitive coupling between the sensor and the first readout stage is needed to avoid electrical stress of the transistors due again to the relatively large bias voltage on the detector side. In addition, the issue can be addressed by using thick oxide devices (with VDD of 2.5 or 3.5 V, to be compared to the typical 1.2 – 1.5 V value of core devices in a 130 nm technology). In order to limit the voltage excursion at the front-end input due to the large pulse signals from the sensor, protection networks might be needed.

In the case of a dual-tier arrangement of the detector, a set of logic circuits performing the coincidence check between the signals from two overlapping pixels needs to be included. The blocks can be integrated and shared between the cells in such a way not to introduce significant discrepancies between the two layers in terms of fill factor. A coincidence signal is generated every time a pixel in the outer layer of the two-tier structure fires within a time window started by the signal from the corresponding pixel in the inner layer.

At an operational gain of $10^6$, an upper limit for the dark count rate R<100 kHz/mm$^2$ can be assumed. With a $\Delta t$ = 10ns coincidence time-window, and N = 400 pixels (pixel size: 50 um x 50 um), the rate of fake 2-fold coincidence is $2 R^2 \Delta t/N$ = 0.5 Hz/mm$^2$. Therefore, for a 1cm$^2$ detector, the rate of accidentals would be close to 50 Hz.

High sensor detector efficiency in the active area of the detector is expected, given the large intrinsic gain of the device and the high probability to trigger an avalanche even in the presence of a small number of ionization charges. The passive structures on the sensor surface do not pose the same problems as in the case of avalanche photo-detectors, given the ability of charged particles to penetrate the surface layer.

Assuming a hit rate of 100 MHz/cm$^2$ and an active area of 1cm$^2$, an average current of 16 uA is expected for a gain of $10^6$, i.e. a power dissipation of about 160 uW/cm$^2$ per layer in the case of a 10V bias for full avalanche operation (readout power excluded). Therefore, for a two-layered sensor, the power dissipation is expected at the level of 0.32 mW/cm$^2$.

Dynamic power consumption is mainly related to the complexity of the readout architecture and to its operating frequency. According to recent estimates [3], relative to a readout network similar to the one we want to implement, the overall power, dominated by the contribution due to the readout network iof n the case of a hit rate 100 MHz/cm$^2$, might be limited to about 0.8 W/cm$^2$.

The distance between the two layers should be less than 20 microns to get a wide semi-aperture angle of 70° with 50 microns pixels. For the two-layered detector thickness this project is aiming to an upper limit of 50 microns (about 0.05% radiation length), taking advantage of the few microns thickness of the depletion region.

As far as the readout electronics is concerned, technologies in the 130 nm node have been extensively characterized and were proven to be extremely tolerant to ionizing radiation. Triplication of the logic blocks in some critical parts of the circuit, together with majority voting architectures may be required in order to harden the system against single event effects.

## 3. Very Preliminary Experimental Proof of the APiX Principle

In order to demonstrate experimentally the validity of the APiX structure detection principle, a very preliminary prototype of a single avalanche pixel structure was assembled starting from a pair of SiPM sensors. Two SiPMs (of 1mm$^2$ area each) are coupled together, positioning the two sensitive areas face-to-face. The active area of the two SiPMs are protected with an epoxy layer and are coupled together so that the distance between them is in the order of tens of microns. This detection structure is a prototype of one avalanche pixel, which is composed of two layers of vertically aligned SiPM avalanche sensors. The signal of each SiPM is read-out through a high bandwidth current amplifier and then is sent to an external logic element. The threshold level is set low enough to produce a logic pulse whenever the signals are produced by the SiPMs, in order to identify the events in which both SiPMs are activated.

The schematic view of the experimental setup is shown in Fig. 3. The photograph (Fig.4) shows the SiPM-based APiX prototype located on the first set of X,Y Silicon strip detectors that were included in the test beam apparatus.

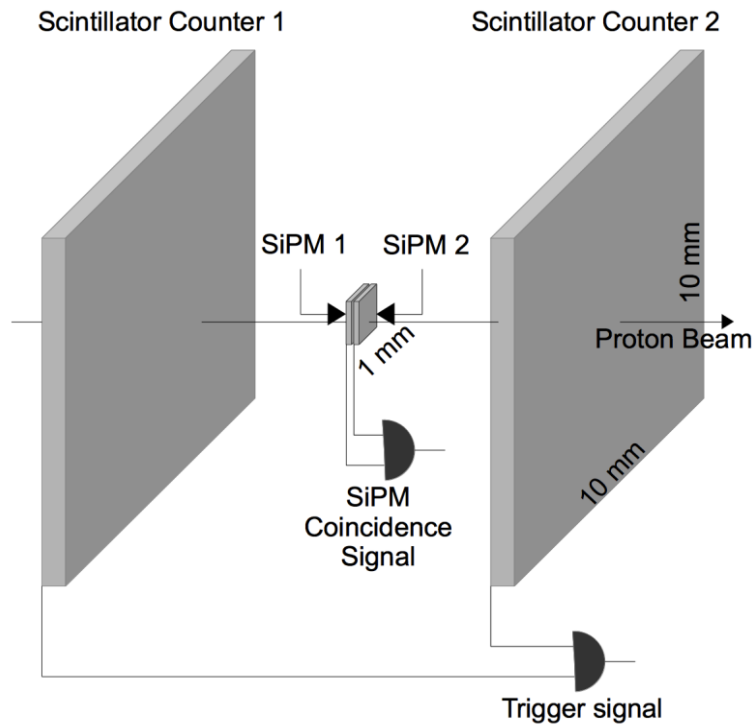

Figure 3. Schematic of the CERN test beam experimental setup

The prototype was exposed to a 120 GeV proton test beam in the CERN North Hall area. This test beam was part of the EUDET E.U. project and the related SiLC R&D [8]. Although two (X,Y) set

– 6 –

of strip Silicon detectors covering 10x10cm$^2$ total area were available in this overall test setup, this very simple first test only used the trigger formed with the signals of two 10x10 cm$^2$ scintillator counters and two 10x10 mm$^2$ monitor counters.

The APiX prototype was placed between the two scintillator counters at a distance of 1 cm. The experiment consists of comparing the trigger rate of the monitor scintillators counters and the coincidence rate of the two SiPMs. The events with two coincident signals in the detectors can be interpreted as detected ionizing particles. According to the geometry of the setup, the ratio between the coincidence SiPM events (Nsipm) and the trigger events (Ntrig) is proportional to the ratio between the sensitive area of the SiPM and the sensitive area of scintillator counters. It is hence expected that $N_{sipm}/N_{trig} = S_{sipm}/S_{trig} \sim 1/10^2$, where S stands for surface ($S_{sipm}$ is the surface covered by the SiPM based prototype and $S_{trig}$ is the one covered by the scintillator based trigger)

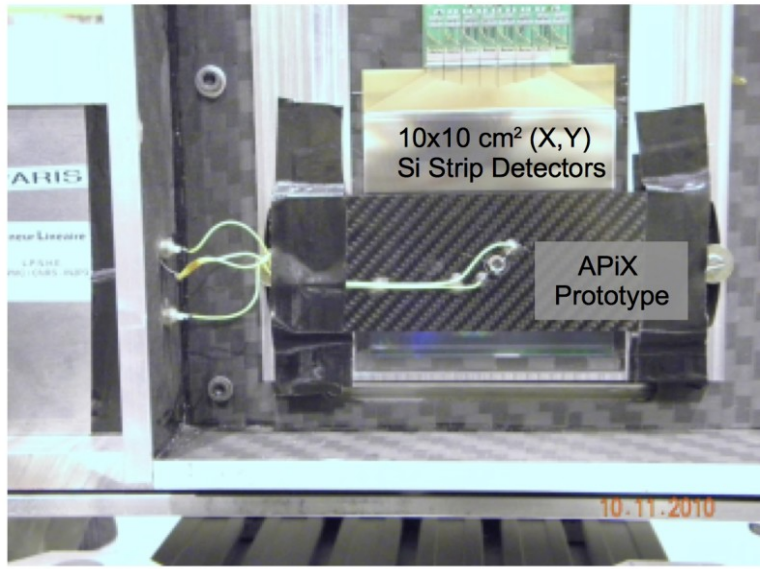

Fig. 4: Test setup of an APiX sensor on the CERN test beam; the APiX prototype is located just in front of a (X,Y) set of 10x10 cm$^2$ Silicon strip detectors

The results of the measurement are reported in Table 1.

| $N_{trig}$ | $N_{SiPMs}$ coinsidence | | $N_{SiPMs}/N_{trig}$ x 10$^{-2}$ |
| --- | --- | --- | --- |
| | measured | corrected | |
| 70320 | 1260 | 1230+-30 | 1.74+-0.05 |
| 103240 | 1283 | 1242+-30 | 1.20+-0.03 |
| 90940 | 1110 | 1070+-30 | 1.18+-0.03 |
| 79725 | 14390 | 14070+-120 | 1.76+-0.02 |

Table 1: Results of the experimental measurements at CERN test beam.

The measurement is performed in four runs. The fluctuations of the different groups of measurements depend mainly on differences in beam conditions. The number of SiPM coincidences is corrected for the random coincidences occurring due to the dark rate. The dark rate of the used SiPMs is about 100 kHz. The logic pulse from each SiPM is 20 ns long. The estimate of uncorrelated dark pulses coincidence rate of the two SiPMs is $2 \times 10^5 \times 10^5 \times 20 \times 10^{-9} = 400$ Hz. The number of measured and corrected coincidences is shown in the second and third column of Table 1. The effect of the dark rate on the measurement is at a level of 2% and does not constitute a problematic contribution.

The average value of the ratio between the SiPM coincidences and the trigger events is measured as $N^{corr}_{sipm}/N_{trig} = (1.47 +- 0.25) \times 10^{-2}$. The alignment of the two SiPMs and the corresponding effective efficiency geometrical contribution is not well controlled in the setup and affects the measurement. Considering this additional systematic uncertainty the result is in reasonable agreement with the expectation of the ratio between the trigger rate and the SiPM rate during the detection of the charged particles as ratio of the sensitive areas.

Improvements are needed in the setup, as the use of a beam pixel telescope for the alignment and as tracking system that will be included for the new APiX demonstrator. This will be the subject of a forthcoming beam test.

## 4. Overview of the Possible Applications

In this section, we briefly outline possible applications of this technology beyond the one we are targeting here and that relates to building a new generation of high precision position detectors for High Energy Physics (HEP) and Astroparticle Physics experiments.

The obvious goal in the former is to use this innovative technology to build new vertex detectors able to confront the challenges of the Super-LHC upgrades and of the future Linear Collider Project with, as timescale, the next decade (i.e. >2020). Although the running conditions have some differences between these two machines and their environmental conditions, they also have a number of similar issues and challenges that this technology should be able to address.

Another aspect is to go to larger size tracking devices, beyond the use as vertex detectors, exploiting the possibility to build pixels of larger size or to design and group them in a flexible way to build different kind of detector geometries.

In addition to the ones mentioned above, there is a large variety of applications requiring high resolution position measurement of charge tracks as, for instance, in mass spectrometers, electron microscopy, Compton spectrometers, particle and isotope identification.

Photon detection can also be addressed with hybrid photo-detectors (like the HPD) whereby the electron kinetic energy of photoelectrons accelerated by a suitable electric field might be sufficient to trigger an APiX cell-pair, opening the way for an high granularity single photon counting device, with low power consumption and low dark noise rate. As an example, an application of this hybrid device for the detection of UV Cherenkov light could be envisaged.

In addition to the several applications in Frontier Research on instrumentation, the technology that will be developed for APiX can impact as well a number of Frontier Industrial Applications. Among those: homeland security, detection of radioactive sources and materials, mobile radiation detection systems, without mentioning the Medical Imaging applications.

## 5. Conclusions

The APiX sensor is a novel coordinate-sensitive ionizing particle detector based on the silicon avalanche structures operated in digital mode with embedded readout electronics on chip. The proposed development of the APiX sensor may open the way to the introduction of an innovative tracking device into the design of the next generation of vertex detectors and tracking systems for High Energy Physics and Astrophysics. The detector concept aims to the minimization of the material thickness traversed by the impinging particle to the benefit of applications where multiple scattering is an issue, by developing a truly thin, digital and 3D device. High rate capability, radiation tolerance, low power, low bias voltage and 3D vertical integration are the main goals of the design. The technological challenge of the vertical integration of aligned sensor-pairs may result into important spin-offs in the domain of frontier research and advanced industrial applications. The detection principle of the APiX detector is experimentally demonstrated at a very preliminary stage and with ionizing radiation.

The design and technology to produce this novel type of ionizing radiation detector is at present under development. This novel type of detector could be interesting for applications of ionizing radiation in many areas, e.g. for position measurement of charged particles in high energy physics experiments, for intensity and position measurement of ionizing radiation in nuclear medicine and others.

**Acknowledgments**